# The distribution dynamics of Carbon Dioxide Emission intensity across Chinese provinces: A weighted Approach

Jian-Xin Wu[1,2] and Ling-Yun He[1,3,*]

[1]School of Economics, Jinan University, China
[2]Business School, University of Western Australia, Australia
[3]School of Economics and Management,
Nanjing University of Information Science and Technology, China
[*]Corresponding Author, Email: lyhe@amss.ac.cn.

**Abstract:** This paper examines the distribution dynamics of carbon dioxide ($CO_2$) emission intensity across 30 Chinese provinces using a weighted distribution dynamics approach. The results show that $CO_2$ emission intensity tends to diverge during the sample period of 1995-2014. However, convergence clubs are found in the ergodic distributions of the full sample and two sub-sample periods. Divergence, polarization and stratification are the dominant characteristics in the distribution dynamics. Weightings with economic and population sizes have important impacts on current distributions and hence long run steady distributions. Neglecting economic size may under-estimate the deterioration in the long run steady state. The result also shows that conditioning on space and income cannot eliminate the multimodality in the long run distribution. However, capital intensity has important impact on the formation of convergence clubs. Our findings have contributions in the understanding of the spatial dynamic behaviours of $CO_2$ emissions across Chinese provinces, and have important policy implications for $CO_2$ emissions reduction in China.

**Acknowledgements**: This paper is supported by the National Social Science Foundation (15ZDA054), Humanities and Social Sciences of Ministry of Education Planning Fund (16YJA790050), and the Natural Science Foundation of China (Nos. 71573258 and 71273261).

# 1. Introduction

Under threat of global warming, $CO_2$ emission abatement has become one of the most important environmental concerns during the past decades. As one of the major energy consumers and $CO_2$ emitters, China has surpassed the United States to be the world largest $CO_2$ emitter since 2007. The huge amount and rapid growth of $CO_2$ emissions in China have attracted considerable attentions from both policy-makers and researchers. Moreover, after decades of rapid economic growth, the income growth is accompanied by serious environmental deterioration. Therefore, the Chinese government is facing the problem of environmental protection and international pressure to reduce $CO_2$ emissions, and have taken actions to reduce $CO_2$ emissions in recent development plans. For example, in the 2009 Copenhagen conference, the Chinese government pledged a $CO_2$ emission intensity (the ratio of $CO_2$ emissions over GDP) reduction of 40%-45% from 2005 to 2020. In 2014, the Government further promised to reach its carbon emissions peak before 2030.

China is a country with a vast territory and 34 heterogeneous provinces of significant diversity. These provinces differ greatly in factor endowment, industry structure, income level and $CO_2$ emissions. More importantly, the national $CO_2$ emission reduction target is disaggregate on the provincial level. Therefore, understanding the distribution dynamics of $CO_2$ emissions across provinces is important in future $CO_2$ emissions prediction and environmental policy making. Therefore, some studies focus on the convergence behaviour of $CO_2$ emissions across provinces in China (Huang and Meng, 2013; Wang and Zhang, 2014; Wang et al., 2014; Hao et al., 2015; Zhao et al., 2015). However, due to different sample periods and econometric approaches, current empirical results remain inherently controversial in existing literature. For example, Wang et al., (2014) find evidence of divergence in $CO_2$ emissions, while many others found evidence of convergence (Huang and Meng, 2013; Wang and Zhang, 2014; Hao et al., 2015; Zhao et al., 2015). Traditionally, there are three types of convergences, namely, sigma, beta, and stochastic convergences. As indicated by Barro and Sala-i-Martin (1992), these traditional approaches are based on models for average or representative economies, which help us understand the evolution of $CO_2$ emissions in general, but provide little information for the spatial distribution dynamics of $CO_2$ emissions across provinces. In addition, these approaches provide no information on the entire shape of the distribution and intradistribution mobility, namely the relative position changes in terms of $CO_2$ emissions. Finally, Chinese provinces differ greatly in economic and population sizes. For example, the economic and population sizes of Guangdong, a southern province in China, are 43 times, and 16 times than those of Ninxia Province, one of Chinese northwest provinces, respectively. The welfare effect of one percent reduction in $CO_2$ emissions in Guangdong is much different from that of Ninxia. To the best of our knowledge, no study yet has yet examined the importance of economic and population sizes in the convergence analysis of $CO_2$ emissions in existing literature.

This paper, therefore, aims at investigating the spatial distribution dynamics of $CO_2$ emission intensity across 30 Chinese provinces (not including Hong Kong, Macau, Taiwan, and Tibet due to data availability). One reason that we choose $CO_2$ emission intensity is that the reduction targets are set based on the $CO_2$ emission intensity in China, rather than per capita $CO_2$ emissions. Hence, the analysis based on $CO_2$ emission intensity has more practical values and policy insights in China. Our analysis differs from those in the existing literature in several aspects: First, we adopt a continuous dynamic distribution approach to examine the dynamic behaviour of $CO_2$ emissions in China. The advantage of this approach is that it can provide dynamic law of entire shape of the distribution for $CO_2$ emissions across provinces. Second, the impact of economic and population sizes are considered in the analysis, which is important in policy-making in China. Third, this paper also explores the factors which may influence the distribution dynamics of $CO_2$ emission intensity in China using a combination of a joint distribution approach and a conditional distribution approach. This further provides information on the formation of convergence clubs.

The remainder of this paper is organized as follows. Section 2 presents a review of the related literature. Section 3 presents the methodology. Section 4 describes the data. Section 5 presents the empirical results and discussions. Section 6 further examines the determinants of the distribution dynamics using a conditional distribution approach. Section 7 provides concluding remarks and discussions of policy implications.

**2. Literature review**
With the increasing concern on the global climate changes, whether $CO_2$ emissions converge or diverge attracts much attention from researchers. However, what is the mechanism behind the convergence in $CO_2$ emissions? Many studies examine the possible existence of an inverted U-shaped relationship known as Environmental Kuznets Curve (EKC) between economic development and environmental pollutions (include $CO_2$ emissions). The rationale behind the EKC is complicated in terms of underlying driving forces. According to Stern (2014), structural change to more informative-intensive and services, improvement in technologies and energy efficiency, increased environmental awareness, and enforcement of environmental regulations, are the forces that may reduce environmental pollutants with the development. Suppose that EKC exists, it's easy to conclude that $CO_2$ emissions would converge over time. In current literature, many theoretical models try to explain the mechanisms behind EKC (Kijima, et al., 2010; Pasten and Figueroa, 2012; Stern, 2014).

The empirical research on convergence of $CO_2$ emissions focuses heavily on both international level across nations and on regional level within a specific country. Strazicich and List (2003) first examine the convergence behaviour of $CO_2$ emissions across countries. They investigate both stochastic and conditional convergences of per capita $CO_2$ emissions across 21 industrial countries using panel unit root tests and cross-section regressions. They find evidence for convergence in the aforesaid countries. Following this work, a growing body of literature examines the convergence of $CO_2$ emissions across countries using different samples and estimation approaches. The first strand of literature uses traditional parametric approaches (Romero-Avila, 2008; Westerlund and Basher, 2008; Barassi et al., 2008; Lee and Chang, 2009; Nourry, 2009; Jobert et al., 2010; Panopoulou and Pantelidis, 2009). Herrerias(2013), Camarero et al., 2013, El-Montasser et al., (2015). This strand of literature examines the existence of absolute, conditional convergence, and stochastic convergence. However, the estimation results are sensitive to the choice of econometric approaches and data sets (Pettersson et al., 2014). Moreover, the traditional approaches only provide limited information on the convergence. For example, β-convergence shows what happens to its mean, while σ-convergence only indicates the dispersion (Sakamoto and Islam, 2008). The second strand of literature adopts nonparametric approaches (Van, 2005; Aldy, 2006; Ezcurra, 2007; Criado and Grether, 2011; Herrerias, 2012; Duro, 2013). Compared with first strand of literature, the second strand of literature not only provides information for the convergence, but also highlights the dynamic law of entire distribution shape for the $CO_2$ emissions across countries, in particular, the existence and formation of convergence clubs.

The huge amount of $CO_2$ emissions in China also attracted great deal of attention in recent years. Huang and Meng (2013) examine the convergence of per capita $CO_2$ emissions in the urban China using a parametric model with spatial-temporal specification. Their results show evidence of convergence for the period 1985-2008. Using the log $t$ test developed by Phillips and Sul (2007), Wang et al. (2014) find evidence for divergence at the country level and three convergence clubs in terms of $CO_2$ emission intensity for the period 1996-2011. Wang and Zhang (2014) study the β-convergence, stochastic convergence and sigma convergence of per capita $CO_2$ emissions in six sectors across 28 provinces in China for the period 1996-2010. Their results indicate convergence in all sectors across the provinces. Hao el al. (2015) examine the convergence of $CO_2$ emission intensity with provincial data for the period 1995-2011. They verified stochastic convergence and β-convergence among Chinese provincial samples. Using a spatial dynamic panel data model, Zhao et al. (2015) examine convergence of $CO_2$ emission intensity across 30 provinces for the period 1990-2010. Their results support the existence of convergence in $CO_2$ emission intensity. Instead of per capita $CO_2$ emissions and $CO_2$ emission intensity, Li and Lin (2015) examine the convergence of energy efficiency with $CO_2$ emissions with provincial panel data.

The conclusions of existing studies on the convergence of $CO_2$ emissions in China are controversially diverse due to the different estimation approaches and samples. As mentioned previously, the traditional convergence approaches have many limitations in estimating the spatial dynamic behaviour of $CO_2$ emission distribution across countries or regions. They provide no information on the intra-distribution dynamics, which may be valuable both theoretically and empirically. For example, multimodality found in the long run steady state may imply the existence of multiple equilibria in the evolution of $CO_2$ emissions. Therefore, instead of single mechanism of the EKC, models of multiple equilibria are expected to shed light on the new stylized facts.

## 3. Research method
*3.1 Weighted distribution dynamics approach*

The distribution dynamics approach is developed by Danny Quah in a series of papers (Quah, 1993, 1996a, 1996b, and 1997) to examine the evolution of income distribution across countries. In comparison to the traditional parametric approaches, the distribution dynamics approach has several advantages. First, it provides more insights on the law of motion in an entire distribution of the variable in question, particularly in the formation of convergence clubs. Second, this approach is completely data-driven, and imposes no assumptions on the model, which helps to avoid the estimation bias due to model assumption in parametric approaches. Third, this approach provides the possibility to account for economic and population sizes in the convergence analysis.

There are two distribution dynamics approaches applied in empirical studies, namely the discrete approach and the continuous approach. To simplify the calculation, most studies use the discrete approach. However, the discrete approach is always criticised by the problem of arbitrarily discretising the variable into different state spaces, which may change the probabilistic properties of the variable in question. In addition, the estimation results of the discrete approach prove to be sensitive to this discretisation process (Quah, 1997; Bulli, 2001; Johnson, 2005). As an upgrading, the continuous dynamic distribution approach employs a stochastic kernel approach to estimate the distribution with infinite state spaces. Therefore, this paper uses the continuous distribution dynamics approach to examine the dynamic behaviour of $CO_2$ emission intensity across Chinese provinces.

Let's suppose that cross province distribution of $CO_2$ emission intensity $x$ at time $t$ can be described by the density function $f_t(x)$. The evolution of the distribution is time-invariant and first-order, that is, the current distribution $f_t(x)$ at time $t$ will evolve into the future distribution $f_{t+\tau}(y)$ at time $t+\tau$, where $\tau>0$. Hence the relationship between the current distribution $f_t(x)$ and the future distribution $f_{t+\tau}(y)$ can be described as follows:

$$f_{t+\tau}(y) = \int_0^\infty g_\tau(y|x) f_t(x) dx \qquad (1)$$

where $g_\tau(y|x)$ is the conditional density function that maps the transition process of the distribution of $CO_2$ emission intensity. Keeping the conditional density function $g_\tau(y|x)$ unchanged, the distribution of $CO_2$ emission intensity will evolve into a long-run equilibrium state, namely, the ergodic distribution. Therefore, the ergodic distribution $f_\infty(y)$ can be estimated as follows:

$$f_\infty(y) = \int_0^\infty g_\tau(y|x) f_\infty(x) dx \qquad (2)$$

To understand the distribution dynamics of $CO_2$ emission intensity, we need to know both the transition dynamics described by $g_\tau(y|x)$ and the long-run equilibrium distribution $f_\infty(y)$. For this purpose, a kernel density approach is used to estimate the conditional density function. The joint natural kernel estimator of $f_{t,t+\tau}(y, x)$ and the marginal kernel estimator of $f_t(x)$ can be defined as follows:

$$f_{t,t+\tau}(y, x) = \frac{1}{n h_x h_y} \sum_{i=1}^n K\left(\frac{x-x_i}{h_x}, \frac{y-y_i}{h_y}\right) \qquad (3)$$

$$f_t(x) = \frac{1}{n h_x} \sum_{i=1}^n K\left(\frac{x-x_i}{h_x}\right) \qquad (4)$$

where $x_i$ is the value of $CO_2$ emission intensity in a specific province at time $t$, and $y_i$ is the value of $CO_2$ emission intensity of that province at $t+\tau$. We use $K(\cdot)$ to denote kernel function. $h_x$ and $h_y$ denote the bandwidth of $x$ and $y$ respectively. In this paper, the bandwidths are estimated using the method by Silverman (1986).

With these definitions, we can further estimate the conditional density with $f_{t,t+\tau}(y,x)$ and $f_t(x)$ through:

$$g_\tau(y|x) = \frac{f_{t,t+\tau}(y,x)}{f_t(x)} \tag{5}$$

In order to account for economic or population sizes, we use population or economic sizes (the share of population /or? GDP) as weights in the analysis. Following Gisbert (2003), the weighted kernel density can be estimated as follows:

$$\hat{f}_t(x) = \frac{1}{nh_x}\sum_{i=1}^{n} \omega_i K_x\left(\frac{x-x_i}{h_x}\right) \tag{6}$$

where $\omega_i$ is the share of GDP or population in province $i$. In this paper, we give out both unweighted and weighted analyses to provide full information of the distributions dynamics of $CO_2$ emission intensity for Chinese provinces.

In the discrete approach, the transition dynamics is presented with transition probability matrix. However, in the continuous approach, three-dimensional and contour plots are the most popular tools to show estimation results of transition probability. The three-dimension plot can show the entire shape of the transition probability distribution, while contour plot has advantages in showing the deviation from the diagonal. Sometimes we need to know the net mobility tendency at each point. To gain more insights into the transition probability mass, we estimate the net transition probability (NTP) values at each point, which is an important index of mobility. The net transition probability index, denoted by $p(x)$, can be defined as:

$$p(x) = \int_x^\infty g_\tau(z|x)dz - \int_0^x g_\tau(z|x)dz \tag{7}$$

According to this definition, a positive net transition probability value at a point indicates an increasing trend in $CO_2$ emission intensity, while a negative net transition probability value at a point implies a decreasing trend. Following common practice in dynamic distribution approaches, we use relative $CO_2$ emission intensity (RCEI), which is the individual province's $CO_2$ emission intensity divided by its yearly average. Therefore, the RCEI values of a specific province correspond to the times of the $CO_2$ emission intensity for this province relative to the provincial average.

*3.2 Joint distribution and conditional distribution approach*

Traditional convergence analysis always examines the convergence clubs with arbitrarily classified regional groups. However, spatial proximity is only a possible reason behind convergence clubs. Other factors, such as income level and capital intensity, may also affect the formation of convergence clubs. Therefore, new approaches are required to account for the formation of convergence clubs. This paper thereby adopts the joint distribution approach and the conditional distribution approach proposed by Quah (1997) to investigate the determinants of the distribution. The joint distribution approach shows the joint kernel density distribution of RCEI and explanatory variables of interest, such as relative income or relative capital. The estimation is similar to that in Equation (3).

In the conditional distribution approach, we need to pre-filter the data to take out the influence of the conditioning variables before the analysis. Specifically, the spatial-conditional RCEI is the RCEI of a province relative to the average of its geographical neighbours. The income and capital intensity conditional RCEI are RCEI divided by relative income and capital intensity respectively. For this reason, the conditional distribution RCEI can be interpreted as the part unexplained by the variable in question. The greater the difference between the distributions of the unconditional and conditional RCEIs, the greater explanatory power of the variable in question.

## 4. Data

There were 34 provincial level administration units in China, namely 22 provinces, 5 autonomous regions, 4 municipalities, and 3 special administration regions (ie., Hong Kong, Macau, and Taiwan). For simplicity, we use the concept of provinces to denote all the provincial administration units. Due to the data availability, we use a panel dataset across 30 provinces (without Hong Kong, Macau, Taiwan, and Tibet) from 1995 to 2014. Our panel data are drawn from official publications of the Chinese statistical agency. The two major sources are the *China Statistical Yearbook* (CSY, various years) and *China Energy Statistical Yearbook* (CESY, various years).

*4.1 $CO_2$ emissions*

The data of $CO_2$ emissions are not directly available in the provincial level in China. As $CO_2$ emissions are generated by the consumption of energy in provinces, we therefore estimate provincial $CO_2$ emissions based on energy consumption. We account for eight primary sources of energy consumptions and $CO_2$ emissions, namely coal, crude oil, coke, gasoline, kerosene, diesel oil, fuel oil, and natural gas. The formula to estimate $CO_2$ emissions is given by:

$$CO_2 = \sum_{i=1}^{n} E_i \times CF_i \times CC_i \times COF_i \times \frac{44}{12} \qquad (8)$$

where *i* indicate the categories of energy. $E_i$, $CF_i$, $CC_i$, and $COF_i$ denote the consumption of energy, the transformation factor, the carbon emission factor, and the carbon oxidation factor, respectively. The estimation is based on the criteria published by the International Panel on Climate Change (IPCC) (2006). Data for energy consumption in each province can be directly obtained from the annual *China Energy Statistical Yearbook*.

*4.2 Population, GDP, and capital*

Our main sources of nominal GDP and population data used in this paper are from the *China Compendium of Statistics 1949-2008*, and the *China Statistic Yearbook* (various years). The capital data are constructed with the same approach as in Wu (2009). The GDP and capital are deflated to the constant price of 1995.

## 5. Empirical results

*5.1 Preliminary analysis*

*5.1.1 The analysis of dispersion*

The coefficient of variation is always used as an indicator of dispersion in convergence analysis. Chinese provinces differ greatly in factor endowment, economic structure, climate, and regional development strategies. Interregional disparity in $CO_2$ emissions is one of the important components of difference in provincial $CO_2$ emissions. Following most studies, we classify Chinese provinces into three regions, namely the eastern, central, and western regions. Figure 1 shows the evolution trend of coefficient of variation and interregional ratios in terms of $CO_2$ emission intensity for the period 1995−2014. We can observe that the coefficient of variation keeps a broadly increasing trend in most years. However, there are two sub-periods, namely 1999-2002, and 2008-2014, when the dispersion in terms of $CO_2$ emission intensity increased sharply. The former is driven by the initiation of the Western Development Program, while the latter is driven by industry relocation policy in recent years. During the sample period, the ratio of central/east maintains a relatively steady state, while the ratio of west/east keeps an increasing trend. Moreover, the evolution trend of ratio of west/east is quite similar to that of coefficient of variation. This may imply that the increase of dispersion is mainly driven by the diversion of $CO_2$ emission intensity in the western region from the other two regions. This further proves that regional preferential policy has significant impact on the spatial distribution of $CO_2$ emissions.

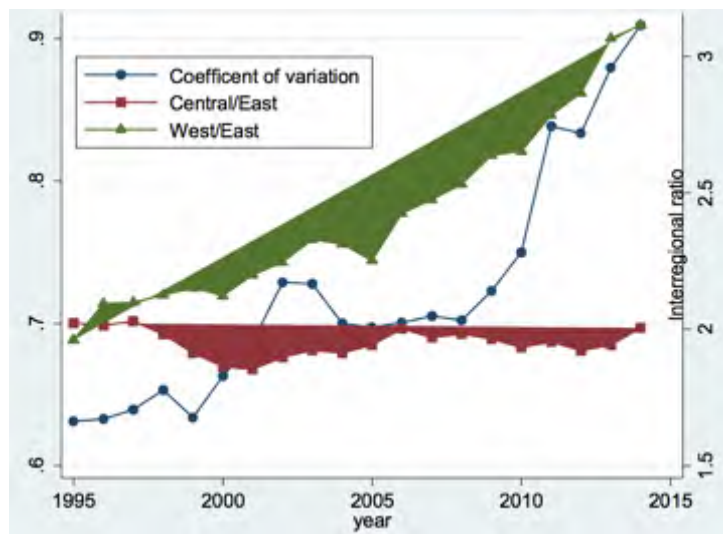

**Figure 1** The coefficient of variation and ratio of the western region and central region to the eastern region in terms of $CO_2$ emission intensity

5.1.2 The analysis on the distributions in critical years

    The coefficient of variation can only provide information on dispersion of $CO_2$ emission intensity across provinces. To get more information about the entire shape of the spatial distribution, Figure 2 plot the unweighted kernel density distribution of RCEI in three representative years, namely, 1995, 2005, and 2014. All three distributions show significant multimodality, with a major peak below mean RCEI, and several small peaks in high RCEI end. Moreover, all three distributions are significantly right-skewed. This indicates that most provinces concentrated in the region below the mean RCEI values, only a small number of provinces locate

in high RCEI end. The major peak in the distribution increases over time in our sample period. However, the distribution of RCEI shrinks slightly at the both ends for the period 1995-2005, while increases significantly at the high end for the period 2005-2014. This suggests divergence of $CO_2$ emission intensity across Chinese provinces, which is consistent with the result of coefficient of variation. Moreover, in comparison to the distribution in 1995, the distribution in 2014 shows more polarization and stratification.

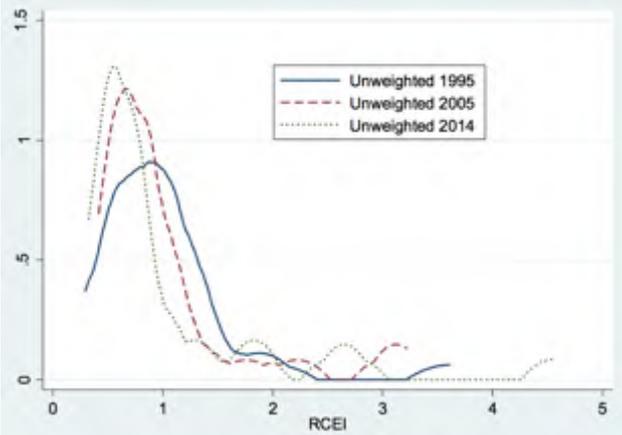

**Figure 2** The kernel density distributions of RCEI in representative years

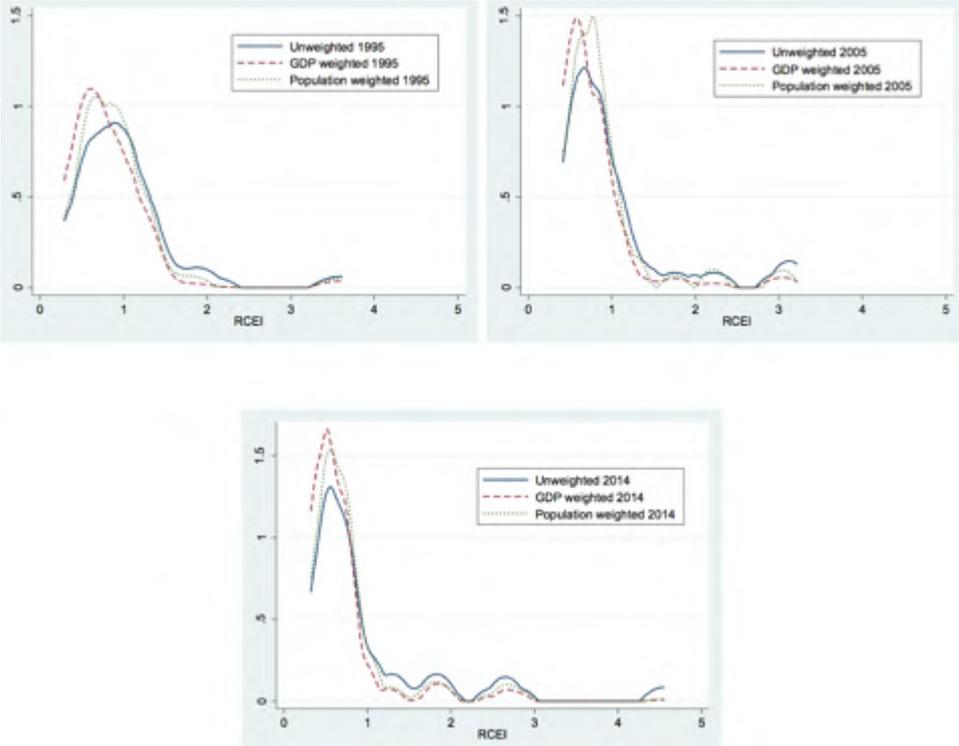

**Figure 3** The weighted distributions of RCEI in 1995, 2005 and 2014

As indicated in previous section, Chinese provinces differ greatly in economic and population sizes, which may have significant impacts on $CO_2$ emissions. Figure 3 shows the economic and population weighted distribution in three representative years. In all the three years' distributions, all economic and population weighted distributions shrink the small peaks in the high RCEI end, but increase the major peak in the low RCEI end. However, weighting with economic size shifts the distribution more to the low RCEI than weighting with population size does. This indicates that provinces with high $CO_2$ emission intensity tend to have smaller economic and population sizes, while provinces with low $CO_2$ emission intensity tend to have larger sizes. Therefore, neglect economic and population sizes may bias the estimation of the real distribution of $CO_2$ emission intensity across Chinese provinces. For example, Ninxia, which is the second smallest provinces in terms of both economic and population (only second to Tibet), has the highest $CO_2$ emission intensity, while Guangdong, which is the largest one in terms of both economic and population sizes, has the lowest $CO_2$ emission intensity.

### 5.2 Distribution dynamics in Chinese provinces

*5.2.1 The full sample*

Figure 4 shows the distribution dynamics of RCEI for the period 1995-2014 with annual transitions. The three-dimensional plot in the panel (a) of Figure 4 shows the distribution of the transitional probability mass with which a region with a specific RCEI value at time *t* could evolve into each value of RCEI at time *t*+1, while the contour plot in the panel (b) of Figure 4 is a top down view of the three-dimensional plot. To interpret this graph, suppose we chose a specific point (RCEI value of 2 for example) on the axis marked *t*, and then slice the plot from this point parallel to axis marked *t*+1; this slice shows the probability distribution of a province with RCEI value of 2 transitioning into each value of RCEI at time *t*+1. In both plots, the concentration of probability mass along the diagonal implies higher persistence and immobility in relative position changes among provinces, while deviation from the diagonal implies higher mobility in relative positions. In comparison with three-dimensional plot, the contour plot shows the deviation from the diagonal line more clearly. Moreover, distinct peaks along the diagonal imply the existence of convergence clubs.

The three-dimensional plot in the panel (a) of Figure 4 shows three peaks along the diagonal, implying the existence of convergence clubs in long run distribution. The contour plot in the panel (b) of Figure 4 shows the deviation of transition probability mass from the diagonal line. It is observed that the transition probability mass is distributed along the diagonal in the low RCEI end, but more deviations from the diagonal line in the high RCEI end. This indicates higher persistence in the provinces with low RCEI values, while relatively high mobility among provinces with high RCEI values.

Net transition probability can provide more accurate information on the convergence of RCEI among provinces. Suppose that low RCEI provinces have positive net transition probability, while high RCEI provinces have negative net transition probabilities, this will imply strong convergence in RCEI. The panel (c) of Figure 4 shows the net transition probability for RCEI of Chinese provinces. Three regions, namely less than RCEI values of 0.85, [1.65, 1.98], and [2.8, 3.5], have positive net transition probability, implying that provinces with RCEI values in these regions have a tendency to increase their $CO_2$ emission intensity. From the policy perspective, these provinces should be imposed with more stringent $CO_2$ reduction targets. Consider that the net transition probability is much higher in the region around RCEI value of 3, the central government should pay more attention on the provinces with three times average $CO_2$ emission intensity.

(a) Three-dimensional plot (unweighted)

(b) Contour map (unweighted)

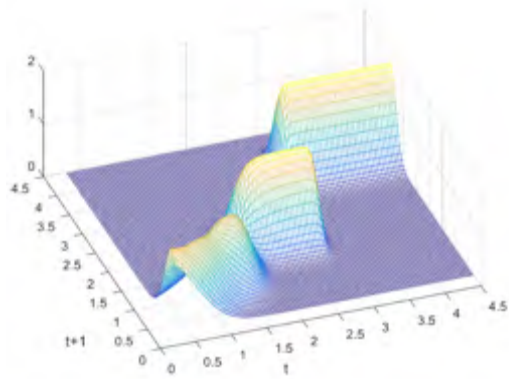
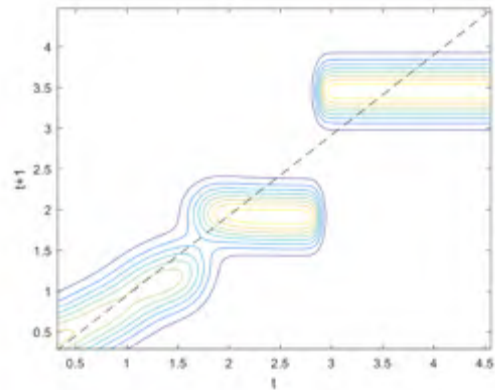

(c) Net transition probability plot (unweighted)

(d) Ergodic distribution (unweighted)

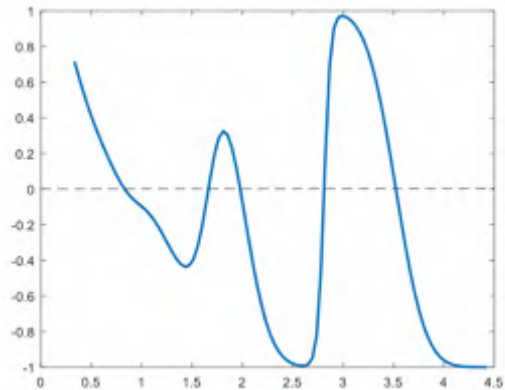
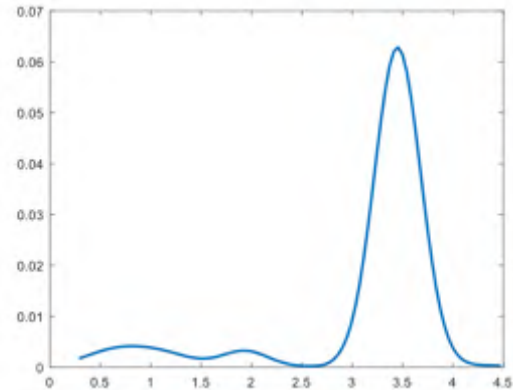

(e) Net transition probability plot (weighted)

(f) Ergodic distribution (weighted)

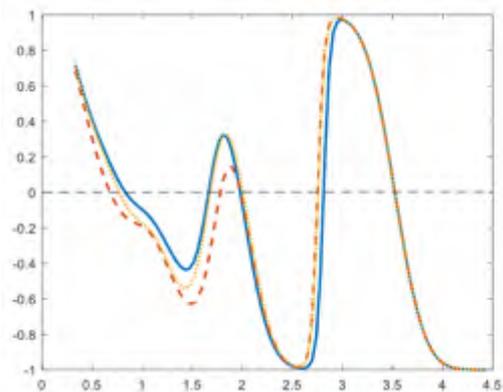
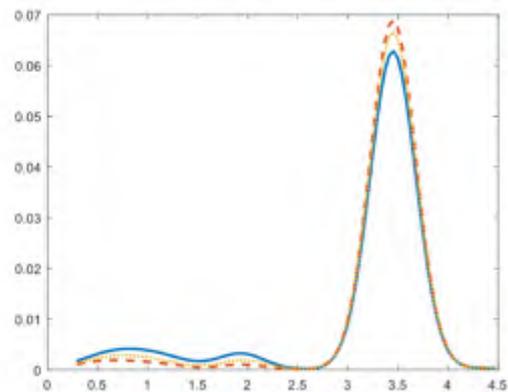

**Figure 4** The distribution dynamics of unweighted and weighted RCEI with annual transition, 1995-2014

*Note*: The solid, dash and dot lines show the unweighted, GDP weighted and population weighted RCEI, respectively.

The panel (d) of Figure 4 presents the ergodic distribution of RCEI based on the transition dynamics of RCEI for the period 1995-2014. Multimodality can be observed in the long run steady distribution, with one major peak around RCEI value of 3.5, and two small peaks around RCEI values of 1 and 2. Keeping the transition dynamics unchanged, Chinese provinces will converge into three clubs differentiated according to their $CO_2$ emission intensity levels in the long run. Moreover, the ergodic distribution is heavily left-skewed, which differ greatly from the current distribution in Figure 2. The ergodic distribution has important implication for policy making: the stratification of provinces in terms of $CO_2$ emission intensity suggest that the government should take policy measures, such as encouraging interregional technology spillover, assigning tougher reduction targets to those provinces clustering into high RCEI clubs, to promote convergence among provinces.

As indicated in previous section, neglecting province heterogeneity in sizes may bias the estimation of welfare effect in some cases. The panels (e) and (f) in Figure 4 show the weighted versions of net transition probability plot and ergodic distribution plot. For comparison, the unweighted curves are also included in the plots. In the panel (e), weighting with economic and population sizes reduces the net transition probability in RCEI values less than 2, but increases net transition probability in the region with RCEI value greater than 2. Moreover, weighting with economic size reduces more net transition probability in low RCEI end than weighting with population size. Correspondingly, the panel (f) of Figure 4 shows that weighting with economic and population sizes does not change the multimodality in the ergodic distribution, but significantly reduces the small peaks in the low RCEI end, while increases the major peak in the high RCEI end. This indicates that some relatively larger provinces tend to converge into high $CO_2$ emission intensity club in the long run.

*5.2.2 Distribution dynamics in two sub-periods*

Before 2005, the Chinese government did not impose tight regulations on $CO_2$ emissions. However, facing with serious environmental deterioration and international pressure, the Chinese government imposed stringent energy consumption and $CO_2$ emission intensity targets in the 11$^{th}$ and 12$^{th}$ Five Year Plans (2005-2010, 2011-2016 respectively). These policy changes may have impact on the distribution dynamics before and after 2005. Therefore, the analysis is further conducted on the two sub-periods, namely 1995-2005, and 2005-2014.

Figure 5 shows the distribution dynamics of RCEI with annul transition for the sub-period 1995-2005. The panels (a) and (b) show the unweighted three-dimensional and contour plots of transition probability mass distribution, while the panels (c) and (d) show the net transition probability and the ergodic distribution plots. Although some small differences can be observed in the shapes of the distributions for transition probability compared with those in Figure 4, the distribution of transition probability mass in Figure 5 is generally similar to that in Figure 4. The ergodic distribution of unweighted RCEI for the period 1995-2005 (solid line in the panel (d) of Figure 5) is significantly multimodality, with two large peaks around the RCEI value of 0.8 and 2, and a small peak around RCEI value of 3.5. The three peaks in the ergodic distribution for the period 1995-2005 are more symmetric than those for the full sample period. Compared with the current distribution for 1995 in Figure 2, a considerable number of provinces have tendency to shift to high $CO_2$ emission intensity end. Moreover, weighting with economic and population sizes significantly increases the peaks around the RCEI value of 0.8 and 3.5, but reduces the peak around the RCEI value of 2. Weighting with economic size has more impacts on the ergodic distribution than weighting with population size does. This implies that Chinese provinces have more disparity in economic size than in population size.

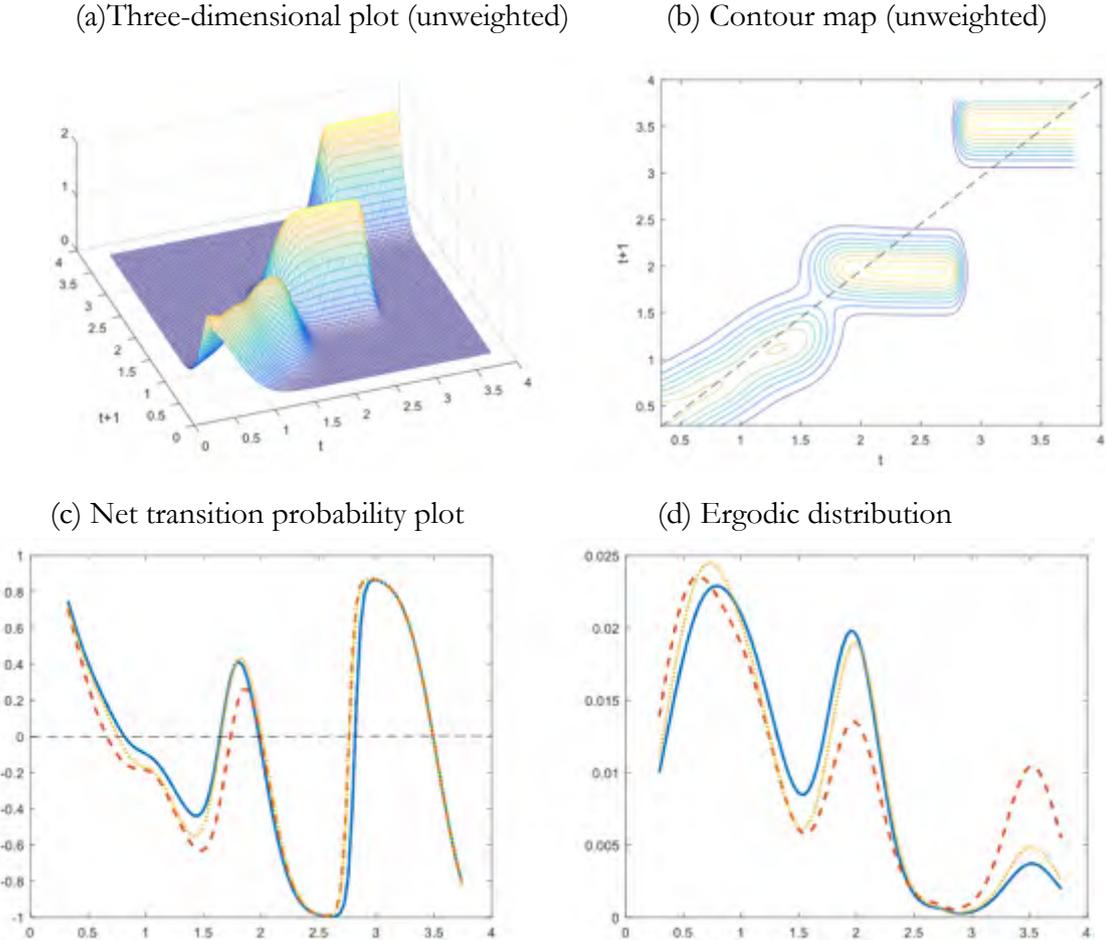

**Figure 5** Net transition probability and ergodic distribution for weighted and unweighted RCEI with annual transition, 1995-2005

*Note*: The solid, dash and dot lines show the unweighted, GDP weighted and population weighted RCEI, respectively.

Figure 6 shows the distribution dynamics for RCEI across Chinese provinces with annual transitions for the sub-period 2005-2014. The distributions of transition probability mass in the panels (a) and (b) in Figure 6 are significantly different from those in Figure 5. This implies that the transitional dynamics of RCEI in the sub-period 1995-2005 is different from that in the sub-period 2005-2014. The net transition probability curve in the panel (c) of Figure 6 also exhibits some differences from that in Figure 5. There is only one region around the RCEI value of 2.8 which has positive net transition probability in the above average region. Significant multimodality, with two small peaks around the RCEI value of 0.8 and 2, and a large peak around the RCEI value of 3.1, can be observed in the long run distribution in the panel (d) of Figure 6. Compared with the right-skewed ergodic distribution in the sub-period 1995-2005, the ergodic distribution in the sub-period 2005-2014 is significantly left-skewed. Keeping the distribution dynamics unchanged in this sub-period, more provinces would converge into the high $CO_2$ emission intensity clubs. In comparison to that in the sub-period 1995-2005, this implies more serious deterioration in the intra-distribution dynamics of RCEI in the sub-period 2005-2014. However, weighting with economic and population sizes has almost opposite effect on the ergodic distribution in this sub-period. Weighting with economic size reduces the peaks in the low RCEI end, but increases the peak in the high RCEI end. On the contrary, weighting with population size increases the peaks in the low RCEI end, but reduces the peak in the high end. This may imply that provinces with larger economy size more tend to shift to high RCEI end than average, while provinces with larger population size have fewer tendencies to shift up.

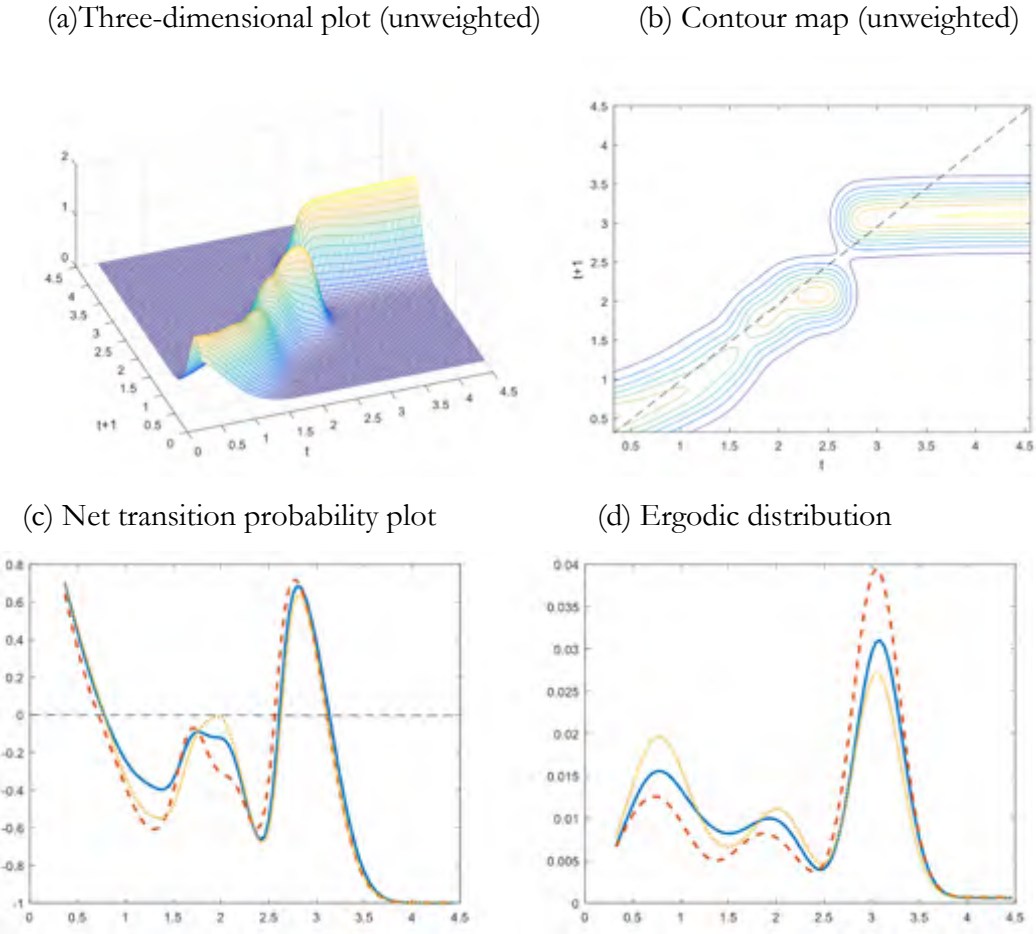

**Figure 6** Net transition probability and ergodic distribution for weighted and unweighted RCEI with annual transition, 2005-2014

*Note*: The solid, dash and dot lines show the unweighted, GDP weighted and population weighted RCEI, respectively.

**6 The determinants of spatial distribution dynamics**

Most studies on the convergence of $CO_2$ emissions only focus on the convergence itself. Few studies try to examine what factors affect the distribution dynamics of $CO_2$ emissions. Different from existing studies, this paper uses a new framework to analyse how geographical location, income level, and capital accumulation influence the distributional dynamics of RCEI with pre-filtered data. For simplicity, we only consider the unweighted analysis in this section.

*6.1 The dynamics of spatial conditional $CO_2$ emission intensity*

The joint distribution approach means that if the RCEI value of a specific province is close to its neighbour's mean value, then the joint probability mass should distribute along the diagonal line. However, the three-dimensional plot (the panel (a) of Figure 7) and contour plot (the panel (b) of Figure 7) show that the joint probability mass deviates significantly from the diagonal line. This indicates that spatial location is not a significant factor in the formation of convergence clubs.

Figure 7 also shows the distribution dynamics of space conditional RCEI with annual transitions for the period 1995-2014. As a comparison, we also include the unconditional curves in the net transitional plot (the panel (e) of Figure 7) and ergodic distribution plot (the panel (f) of Figure 7). The three-dimensional plot (the panel (c) of Figure 7), contour plot (the panel (d) of Figure 7) and net transitional probability plot (the panel (e) of Figure 7) show that the transition dynamics of the space conditional RCEI exhibits some differences than those in the unconditional case in Figure 4. More importantly, multimodality still exists in the ergodic distribution of space conditional RCEI. Conditioning on neighbour's mean value does not eliminate convergence clubs in the long run steady state. The distribution of conditional RCEI can be considered as the part unexplained by the variable in question. Hence, the above result indicates that the convergence clubs in the ergodic distribution of RCEI cannot be explained by geographical proximity. This result differs from that in Wang et al. (2014), which find the existence of regional convergence clubs with parametric approach. However, the parametric approach only provides the statistic result of 'representative economy'. Our result is more intuitive with more details than those in previous studies.

*6.2 The dynamics of income conditional $CO_2$ emission intensity*

As discussed previously, income level is an important factor that may influence $CO_2$ emissions. The inverted U-shaped relationship between $CO_2$ emissions and income level, known as 'environmental Kuznets curve' (EKC), has been well documented in many studies. However, some studies (Bassetti et al., 2013) indicate that the nexus of income level and $CO_2$ emissions may be more complex than that predicted by EKC paradigm. For convenience, this paper employs relative income (RI) that normalized to the yearly average in the analysis.

The panels (a) and (b) in Figure 8 show the three-dimensional and contour plots of the joint distribution between RCEI and RI. It is observed that the probability mass distributes along two axes. Although most provinces locate in the highest peak in low income and low $CO_2$ emission intensity region, there are still considerable number of provinces that concentrate in the peaks along two axes. 'Low income, high pollution' provinces co-exist with 'high income, low pollution' ones in China. This phenomenon cannot be explained by the EKC theory.

(a) Three-dimensional plot of joint distribution      (b) Contour plot of joint distribution

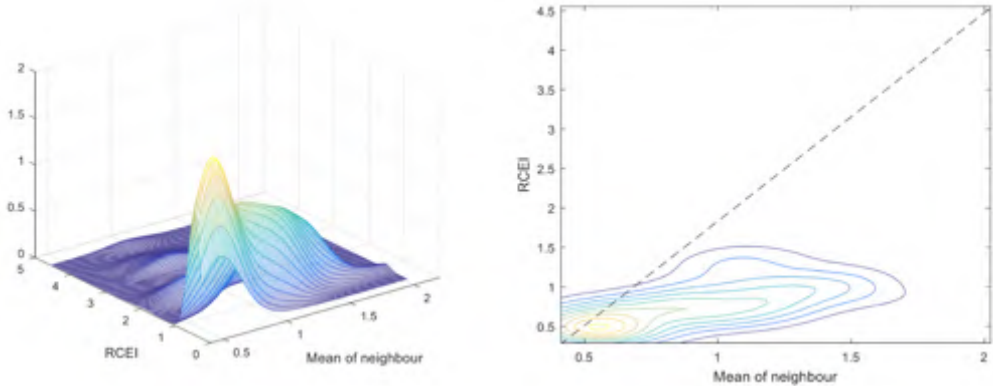

(c) Three-dimensional plot                          (d) Contour plot

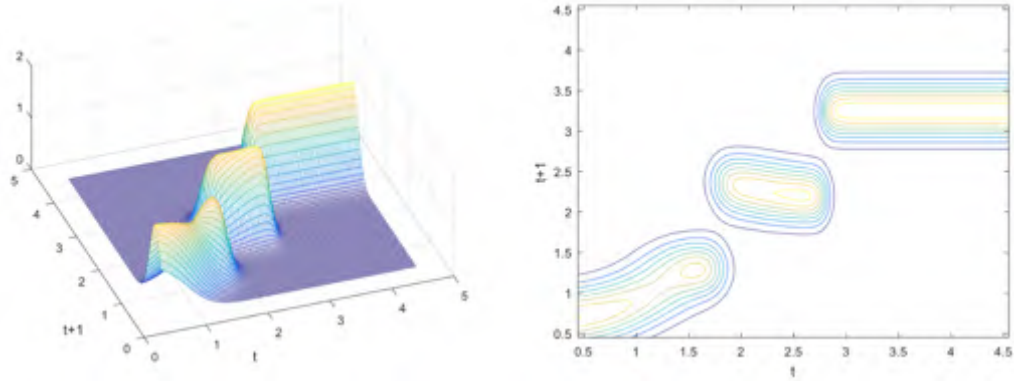

(e) Net transition probability plot             (f) Ergodic distribution

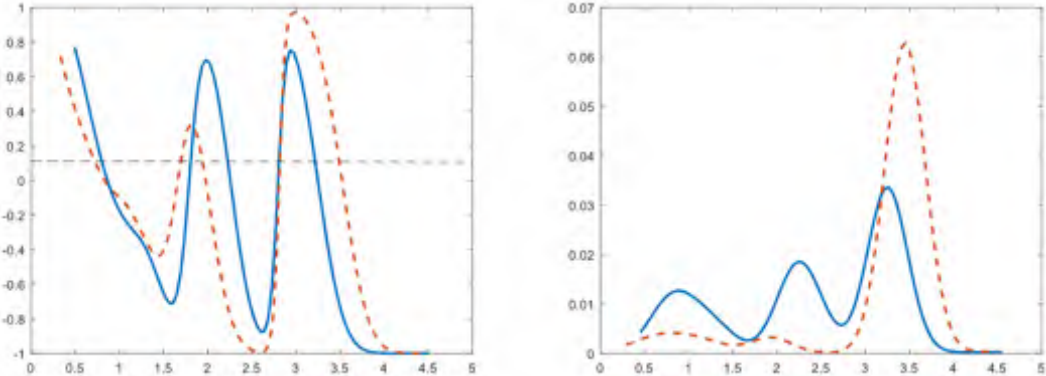

**Figure 7** The distribution dynamics of the space conditional RPCG with annual transition, 1978-2013

*Note*: The solid and dash lines show the conditional and unconditional distributions, respectively.

The panels (c) and (d) in Figure 8 show the distributions of transition probability mass for income conditional RCEI with annual transitions for the period 1995-2014. There are some differences in the distribution of transition probability between the income-conditional RCEI in Figure 8 and unconditional one in Figure 4. The net transition probability curve of conditional RCEI (solid line in the panel (e) of Figure 8) also differs from that of unconditional RCEI (dash line in the panel (e) of Figure 8). Conditioning on income show that income may have some impact on the distribution dynamics of $CO_2$ emission intensity. However, the multimodality can still be observed in the ergodic distribution of income conditional RCEI (solid line in panel (f) of Figure 8). Conditioning on income cannot eliminate convergence clubs observed in the unconditional distribution.

(a)Three-dimensional plot of joint distribution   (b) Contour plot of joint distribution

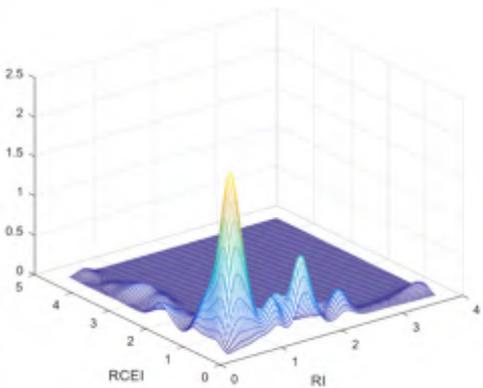
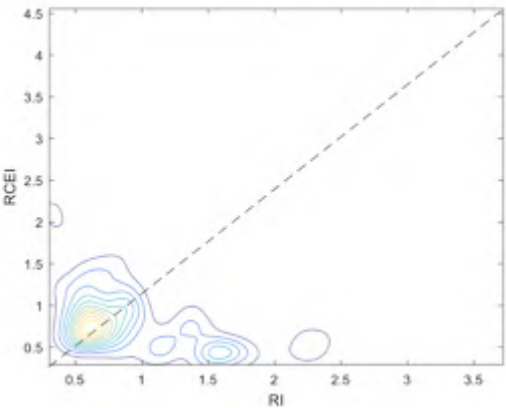

(c)Three-dimensional plot   (d) Contour plot

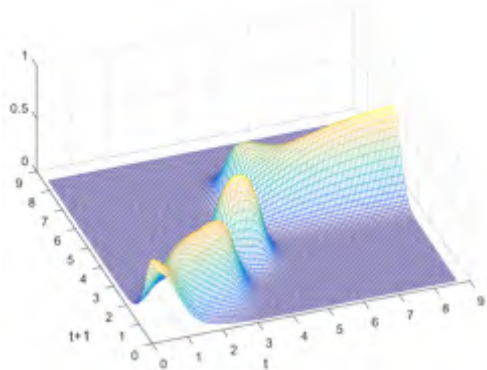
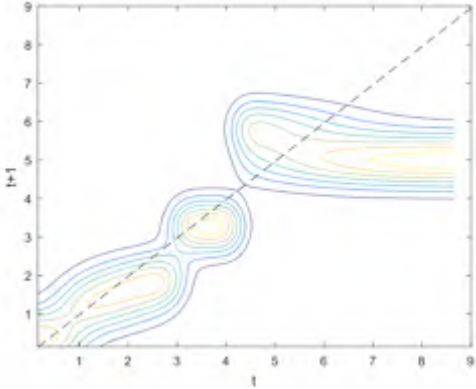

(e) Net transition probability plot   (f) Ergodic distribution

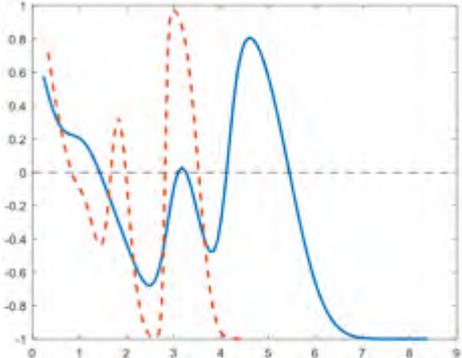
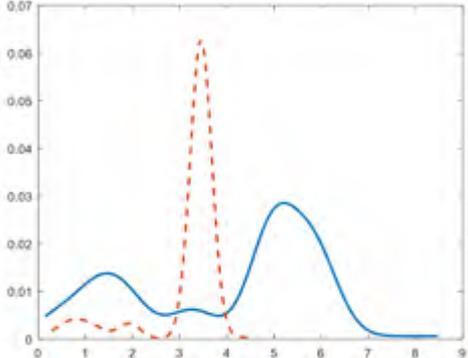

**Figure 8** The distribution dynamics of the income conditional RCEI with annual transition, 1995-2014

*6.3 The distribution dynamics of capital conditional CO$_2$ emission intensity*

(a) Three-dimensional plot of joint distribution       (b) Contour plot of joint distribution

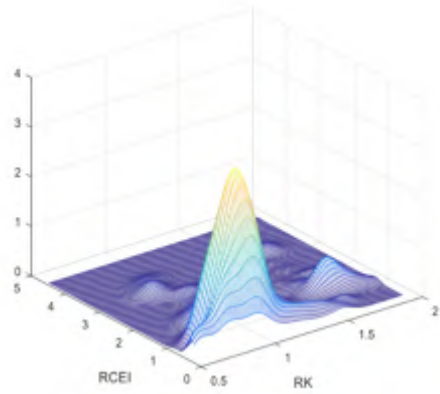
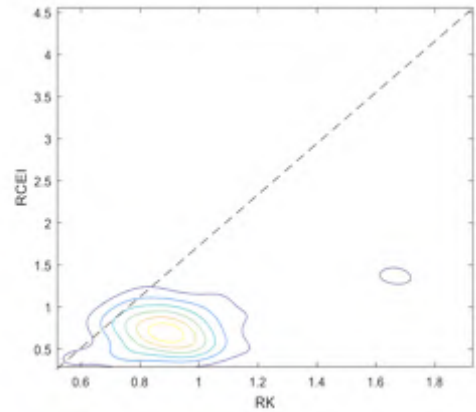

(c) Three-dimensional plot                                            (d) Contour plot

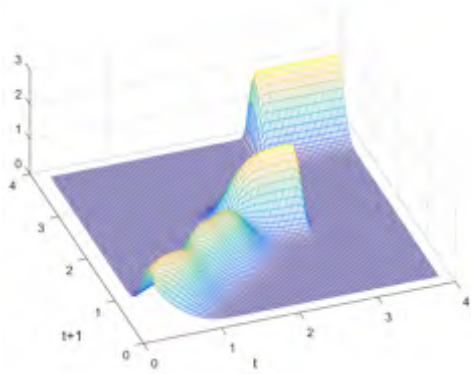
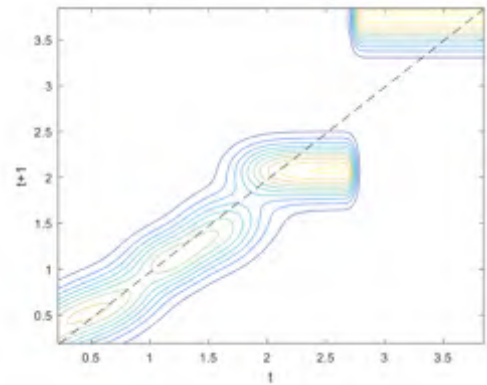

(e) Net transition probability plot                                  (f) Ergodic distribution

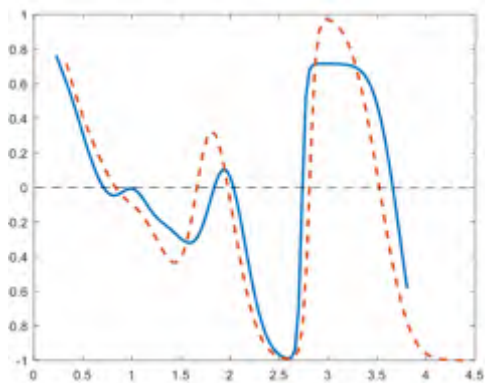
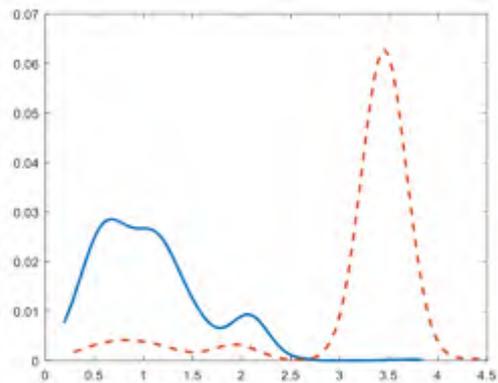

**Figure 9** The distribution dynamics of the capital conditional RCEI with annual transition, 1995-2014

*Note*: The solid and dash lines show the conditional and unconditional distributions, respectively.

High capital intensity is always related with high energy consumption and hence high CO$_2$ emissions. In this subsection, we focus on the relationship between CO$_2$ emission intensity and capital intensity using joint distribution approach and conditional approach. Panel (a) and panel (b) in Figure 9 show the joint distribution of RCEI and relative capital (RK). As most of the density concentrated in the below average area, no simple relationship can be observed in the joint distribution.

Panel (c) and panel (d) in Figure 9 show the distribution of transition probability mass for capital conditional RCEI with annual transitions for the period 1995-2014. In Comparison with the unconditional case in Figure 4, the distribution dynamics of capital conditional RCEI differs from the unconditional one. The ergodic distribution for the conditional RCEI is bimodality, which is different from multimodality of unconditional RCEI. The highest peak in the ergodic distribution disappears and the density mass are concentrated around the average. This indicates that capital intensity is an important factor in the formation of convergence clubs, the high $CO_2$ emission intensity club in particular. These results have important policy implication. In China, government play important role in capital investment. Hence, the government can promote convergence in $CO_2$ emission intensity through capital investment policy.

## 7. Conclusion and policy implications

This paper examines the distribution dynamics of $CO_2$ emission intensity across 30 provinces for the period 1995-2014. Different from existing studies, this paper adopts a weighted distribution dynamics approach accounting for both economic and population sizes. Moreover, we use a combination of joint distribution approach and a conditional distribution approach to examine the determinants of the distribution dynamics of $CO_2$ emission intensity.

The result shows that Chinese provincial $CO_2$ emission intensity has a dispersion trend in general. This dispersion is mainly driven by the increase of $CO_2$ emission intensity in the western region relative to the eastern and central regions. The paper finds more persistence in provinces with low $CO_2$ emission intensity and more mobility in provinces with high $CO_2$ emission intensity. Convergence clubs can be observed in the long run steady state. The intradistribution dynamics shows that most provinces converge into high $CO_2$ emission intensity end. This indicates the deterioration in the evolution trend of $CO_2$ emission intensity. In general, different from existing studies (such as Huang and Meng (2013), Wang and Zhang(2014), Wang et al.(2014), Hao et al.(2015), Zhao et al.(2015)), our findings show that divergence, polarization and stratification are the dominant characteristics in the distribution dynamics of $CO_2$ emission intensity across Chinese provinces. Weighting with economic and population sizes have significant impacts on the distribution dynamics of $CO_2$ emission intensity. It increases the peak in the high $CO_2$ emission intensity end, but reduces the peak in the low $CO_2$ emission intensity end. Neglecting economic and population sizes may underestimate the deterioration in the long run steady state. The analysis on the two sub-periods shows that the deterioration is more significant in the sub-period 2005-2014 than that in the sub-period 1995-2005. The conditional analysis indicates that conditioning on space and income cannot eliminate the multimodality in the ergodic distribution, implying that the convergence clubs in the long run steady state is not determined by space and income in general. However, the result shows that capital intensity have significant impacts on the formation of convergence club in the high $CO_2$ emission intensity end.

The results of this paper have important policy implications. First, the lack of convergence and existence of convergence clubs in the distribution of $CO_2$ emission intensity shows that the market cannot reduce environmental pollutants automatically. Policies focused on technological spillover and industry upgrading are required to promote convergence in $CO_2$ emission intensity. For example, Ninxia, which has the highest $CO_2$ emission intensity in recent years, has a large share of heavy industries, such as coal mining industry, metallurgical industry and mechanical industry. Therefore, technological progress and industry upgrading are the most efficient way to reduce $CO_2$ emission intensity. Second, in China, $CO_2$ emission reduction targets are assigned in the provincial level. However, provinces differ greatly in economic and population sizes. Hence, economic and population sizes should be considered in policy-making. Considering the deterioration in the intradistribution of $CO_2$ emission intensity, the central government should assign tougher targets to the provinces which have tendency to increase their relative $CO_2$ emission intensity, particularly those with high $CO_2$ emission intensity.

**Reference**
Aldy, J. E., 2006. Per capita carbon dioxide emissions: convergence or divergence? Environmental and Resource Economics, 33(4), 533-555.
Barassi, M. R., et al., 2008. Stochastic divergence or convergence of per capita carbon dioxide emissions: re-examining the evidence. Environmental and Resource Economics 40(1), 121-137.
Barro, R. J., Sala-i-Martin, X., 1992. Convergence. Journal of political Economy, 223-251.
Bassetti, T., et al., 2013. $CO_2$ emissions and income dynamics: What does the global evidence tell us?. Environmental and Resource Economics, 54(1), 101-125.
Bulli, S., 2001. Distribution dynamics and cross-country convergence: a new approach. Scottish Journal of Political Economy, 48(2), 226-243.
Camarero, M., et al., 2013. Are the determinants of CO 2 emissions converging among OECD countries?. Economics Letters, 118(1), 159-162.
CCS, 2010. China Compendium of Statistics 1949-2008. Beijing: China Statistics Press.
CESY (various years). China Energy Statistical Yearbook. Beijing: China Statistics Press.
Criado, C. O., Grether J.-M., 2011. Convergence in per capita $CO_2$ emissions: a robust distributional approach. Resource and Energy Economics, 33(3), 637-665.
CSY (various years). China Statistical Yearbook. Beijing: China Statistics Press.
Duro, J. A., 2013. International mobility in carbon dioxide emissions. Energy policy, 55, 208-216.
El-Montasser, et al., 2015. Convergence of greenhouse gas emissions among G7 countries. Applied Economics, 47(60), 6543-6552.
Ezcurra, R., 2007. Is there cross-country convergence in carbon dioxide emissions? Energy Policy, 35(2), 1363-1372.
Gisbert, F. J. G., 2003. Weighted samples, kernel density estimators and convergence. Empirical Economics, 28(2), 335-351.
Hao, Y., et al., 2015. Is China's carbon reduction target allocation reasonable? An analysis based on carbon intensity convergence. Applied Energy, 142, 229-239.
Herrerias, M. J., 2012. $CO_2$ weighted convergence across the EU-25 countries (1920–2007). Applied energy, 92, 9-16.
Herrerias, M. J., 2013. The environmental convergence hypothesis: Carbon dioxide emissions according to the source of energy. Energy policy, 61, 1140-1150.

Huang, B., Meng L., 2013. Convergence of per capita carbon dioxide emissions in urban China: A spatio-temporal perspective. Applied Geography, 40, 21-29.

Jobert, T., et al., 2010. Convergence of per capita carbon dioxide emissions in the EU: Legend or reality? Energy Economics, 32(6), 1364-1373.

Johnson, P. A., 2005. A continuous state space approach to "convergence by parts". Economics Letters, 86(3), 317-321.

Kijima, M., et al., 2010. Economic models for the environmental Kuznets curve: A survey. Journal of Economic Dynamics and Control, 34(7), 1187-1201.

Lee, C. C., Chang, C. P., 2009. Stochastic convergence of per capita carbon dioxide emissions and multiple structural breaks in OECD countries. Economic Modelling, 26(6), 1375-1381.

Li, K., Lin B., 2015. Metafroniter energy efficiency with $CO_2$ emissions and its convergence analysis for China. Energy Economics, 48, 230-241.

Nourry, M., 2009. Re-examining the empirical evidence for stochastic convergence of two air pollutants with a pair-wise approach. Environmental and Resource Economics, 44(4), 555-570.

Panopoulou, E., Pantelidis T., 2009. Club Convergence in Carbon Dioxide Emissions. Environmental and Resource Economics, 44(1), 47-70.

Pasten, R., Figueroa, E., 2012. The environmental Kuznets curve: a survey of the theoretical literature. International Review of Environmental and Resource Economics, 6(3), 195-224.

Pettersson, F., et al., 2014. Convergence of carbon dioxide emissions: a review of the literature. International Review of Environmental and Resource Economics, 7(2),141-178.

Phillips, P. C., Sul, D., 2007. Transition modeling and econometric convergence tests. Econometrica, 75(6), 1771-1855.

Quah, D. T., 1993. Empirical cross-section dynamics in economic growth. European Economic Review, 37(2), 426-434.

Quah, D. T., 1996a. Empirics for economic growth and convergence. European Economic Review, 40(6), 1353-1375.

Quah, D. T., 1996b. Twin peaks: growth and convergence in models of distribution dynamics. The Economic Journal, 106(437), 1045-1055.

Quah, D. T., 1997. Empirics for growth and distribution: stratification, polarization, and convergence clubs. Journal of economic growth, 2(1), 27-59.

Romero-Ávila, D., 2008. Convergence in carbon dioxide emissions among industrialised countries revisited. Energy Economics, 30(5), 2265-2282.

Sakamoto, H., Islam, N., 2008. Convergence across Chinese provinces: an analysis using Markov transition matrix. China Economic Review, 19(1), 66-79.

Silverman, B. W., 1986. Density estimation for statistics and data analysis, CRC press.

Stern, D. I., 2014. The environmental Kuznets curve: A primer. Centre for Climate Economics & Policy, Crawford School of Public Policy, The Australian National University.

Strazicich, M. C., List J. A., 2003. Are $CO_2$ emission levels converging among industrial countries? Environmental and Resource Economics, 24(3), 263-271.

Van, P. N., 2005. Distribution dynamics of $CO_2$ emissions. Environmental and Resource Economics, 32(4), 495-508.

Wang, J., Zhang K., 2014. Convergence of carbon dioxide emissions in different sectors in China.

Energy, 65, 605-611.

Wang, Y., et al., 2014. Convergence behavior of carbon dioxide emissions in China. Economic Modelling, 43, 75-80.

Westerlund, J., Basher S. A., 2008. Testing for convergence in carbon dioxide emissions using a century of panel data. Environmental and Resource Economics 40(1), 109-120.

Wu, Y. 2009. China's capital stock series by region and sector. The University of Western Australia Discussion Paper 09.02.

Zhao, X., et al, 2015. Province-level convergence of China's carbon dioxide emissions. Applied Energy, 150, 286-295.